\documentclass[twocolumn]{aastex6}

\usepackage{graphicx}

\begin{document}

\shortauthors{Ribas et al. 2018}
\shorttitle{The VLA view of HD 98800}

\title{Long-lived protoplanetary disks in multiple systems: the
  VLA view of HD 98800}

\author{\'Alvaro Ribas\altaffilmark{1}, 
Enrique Mac\'ias\altaffilmark{1},
Catherine C. Espaillat\altaffilmark{1},
Gaspard Duch\^ene\altaffilmark{2,3}}
\altaffiltext{1}{Department of Astronomy, Boston University, Boston, MA 02215, USA; {\tt aribas@bu.edu}}
\altaffiltext{2}{Department of Astronomy, UC Berkeley, Berkeley, CA 94720, USA}
\altaffiltext{3}{Univ. Grenoble Alpes/CNRS, IPAG, F-38000 Grenoble, France}

%%%% ABSTRACT
\begin{abstract}

  The conditions and evolution of protoplanetary disks in multiple systems can be
  considerably different from those around single stars, which may have important
  consequences for planet formation. We present Very Large Array (VLA) 8.8\,mm (34\,GHz)
  and 5\,cm (6\,GHz) observations of the quadruple system HD~98800, which consists of two
  spectroscopic binary systems (Aa-Ab, Ba-Bb). The Ba-Bb pair is surrounded by a
  circumbinary disk, which is usually assumed to be a debris disk given its $\sim$10\,Myr age and
  the lack of near infrared excess. The VLA 8.8\,mm observations resolve the disk size
  (5-5.5\,au) and its inner cavity ($\approx$3\,au) for the first time, making it one of
  the smallest disks known. Its small size, large fractional luminosity, and millimeter
  spectral index consistent with blackbody emission support the idea that HD~98800~B is a
  massive, optically thick ring that may still retain significant amounts of gas. The
  disk detection at 5\,cm is compatible with free-free emission from photoionized
  material. The diskless HD~98800~A component is also detected, showing partial
  polarization at 5\,cm that is compatible with nonthermal chromospheric activity. We propose
  that tidal torques from Ba-Bb and A-B have stopped the viscous evolution of the inner
  and outer disk radii, and the disk is evolving via mass loss through photoevaporative
  winds. This scenario can explain the properties and longevity of HD~98800~B as well as
  the lack of a disk around HD~98800~A, suggesting that planet formation could have more
  time to proceed in multiple systems than around single stars in certain system
  configurations.
\end{abstract}

\keywords{binaries: general --- protoplanetary disks --- stars: individual
(HD~98800) --- stars: pre-main sequence --- techniques: interferometric}

%%%% INTRODUCTION
\section{Introduction} \label{sec:intro}

Given the abundance of binaries and multiple systems in our Galaxy, understanding planet
formation in these systems represents an important piece of our knowledge of exoplanetary
populations. In fact, 30-75\,\% of stars appear to have formed in multiple systems
\citep{Duquennoy1991, Kraus2008, Lafreniere2008, Kraus2011, Duchene2013}, and various
exoplanets have already been found around binary and multiple systems
\citep[e.g.][]{Doyle2011, Welsh2012_kepler_binaries, Dupuy2016, Kostov2016}.

The formation of planets around single stars occurs in the protoplanetary disks
surrounding young stars. These disks contain gas and dust which, via dust growth first and
gas accretion later, can form rocky and gas giant planets \citep{Raymond2014, Helled2014,
  Testi2014}. At the same time, processes such as the viscous evolution of the disk
\citep[e.g.][]{Shakura1973, Lynden-Bell1974, Hartmann1998}, photoevaporation by stellar
radiation \citep[e.g.][]{Shu1993, Hollenbach1994}, or the interaction with newborn planets
\citep[][]{Kley2012, Espaillat2014} lead to the eventual dispersal of the disk
\citep[see][and references therein]{Alexander2014}, which occurs 5-10\,Myr after their
formation \citep{Haisch2001, Hernandez2007, Hernandez2008, Mamajek2009, Ribas2015}. Once
the protoplanetary disk disperses, a second generation of dust produced by planetesimal
collisions (with some possible contribution from remaining dust from the protoplanetary
phase) forms a gas-poor, less massive, colder disk called a debris disk, analogous to our
own Kuiper Belt \citep[e.g.][]{Wyatt2008, Hughes2018}. While the latter can survive for
much longer ($\sim$Gyr) thanks to dust replenishment from these collisions, the transition
from a protoplanetary to a debris disk is still not well understood. The importance of
disk evolution and dispersal is obvious, since it sets a time limit to planet formation
and determines many of the properties of the resulting planetary systems.

This scenario can change significantly in binary or multiple systems; the gravitational
interaction of a (sub)stellar companion can have an important effect on disks, leading to
truncation or even quick dispersal depending on various parameters, such as the semi-major
axis of the binary orbit, its eccentricity, or the masses of the components
\citep[e.g.][]{Papaloizou1977, Artymowicz1994}. In fact, various surveys have shown the
fraction of protoplanetary disks to be smaller around close ($a< 40$\,au) binaries than
around single stars or wide binaries \citep{Cieza2009, Harris2012, Kraus2012}. On the
other hand, very close binaries (a few au) may not be able to retain individual
circumstellar disks (disks around individual components of the system), but could harbor a
circumbinary disk surrounding both stars. Mass accretion rates in circumbinary disks can be
severely diminished due to the tidal torques exerted by the central sources, resulting in
extended disk lifetimes \citep[e.g.][]{Alexander2012}. Also, disks with truncated outer
radii due to a close companion may even experience an outside-in evolution
\citep{Rosotti2018}, which is completely different from the expectation for disks around single
stars. All these mechanisms are likely to result in different populations of exoplanets
around binary and multiple stars and are thus key pieces for understanding them.

In this study, we present NSF's Karl G. Jansky Very Large Array (VLA) observations of an
extreme case of a circumstellar disk in a multiple system: the hierarchical quadruple
system HD~98800. The VLA observations resolved both the size and the inner cavity of the
disk around HD~98800~B for the first time and shed light onto the puzzling existence of
long-lived disks in multiple systems. The paper is outlined as follows:
Sec.~\ref{sec:description} provides a description of the HD~98800
system. Sec.~\ref{sec:observations} describes the VLA observations and ancillary data
used, and the results and modeling procedure are discussed in Sec.~\ref{sec:results}. In
Sec.~\ref{sec:discussion}, we examine the nature of the disk around HD~98800~B, compare
our results with previous studies, and discuss implications for disks in multiple
systems. Finally, our conclusions are presented in Sec.~\ref{sec:conclusions}.

%%%% DESCRIPTION
\section{The HD~98800 system}\label{sec:description}

HD~98800 is a hierarchical quadruple system located in the 7-10\,Myr old TW~Hya
association \citep{Soderblom1996, Webb1999, Prato2001, Torres2008, Ducourant2014}. It
  was first identified as a possible T~Tauri star by \citet{Gregorio-Hetem1992}, together
  with other members of TW~Hya. At $\approx$\,45\,pc \citep{vanLeuven2007}, this source
comprises two spectroscopic binary systems (A and B) orbiting each other with a semi-major
axis of $\approx$\,45\,au \citep[P = 214\,years,][]{Tokovinin2014}. While the orbital
parameters of the southern component (Aa-Ab) are uncertain, \citet{Boden2005} used
interferometric data to measure the orbits of the northern one (Ba-Bb) and found two stars
of comparable masses ($M_{\rm Ba}=0.7$\,$M_\odot$, $M_{\rm Bb}=0.6$\,$M_\odot$) orbiting with a
semi-major axis of $a\approx0.5$\,au (P = 315\,days) and a high eccentricity
($e=0.78$). The orbital period of Aa-Ab is believed to be similar \citep{Torres1995}. The
relevant properties of HD~98800 adopted in the paper are listed in
Table.~\ref{tab:HD98800}.

\begin{table}
  \caption{Properties of HD~98800.}\label{tab:HD98800}
  \begin{center}
    \begin{tabular}{l c c }
      \hline
      \hline\rule{0mm}{3mm}Parameter & Value & Reference \\
      \hline
      Distance & 45\,pc & 1 \\
      Age & 7-10\,Myr & 2 \\
      \hline
      \multicolumn{3}{c}{A Component (Aa+Ab)} \\
      \hline
      $a_{\rm A-B}$ & 45\,au & 3\\
      $e_{\rm A-B}$ & 0.4 & 3\\
      $i_{\rm A-B}$ & 88$^{\circ}$ & 3\\     
      $M_{\rm A}$(Aa+Ab)$^\dagger$ & 1.3\,$M_\odot$& 4\\
      $A_v$ & 0\,mag & 5\\
      \hline
      \multicolumn{3}{c}{B Component} \\
      \hline
      $a_{\rm Ba-Bb}$ & 0.5\,au & 6\\
      $e_{\rm Ba-Bb}$ & 0.78 & 6\\
      $i_{\rm Ba-Bb}$ & 67$^{\circ}$ & 6\\
      $M_{\rm Ba}$ & 0.7\,$M_\odot $& 6\\
      $M_{\rm Bb}$ & 0.6\,$M_\odot $& 6\\
      $T_{\rm eff, Ba}$ & 4200\,K & 6\\
      $T_{\rm eff, Bb}$ & 4000\,K& 6\\
      $R_{\rm eff, Ba}$ & 1.1\,$R_\odot $& 6\\
      $R_{\rm eff, Bb}$ & 0.9\,$R_\odot $& 6\\
      $A_v$ & 0.44\,mag & 7\\
      \hline
    \end{tabular}
  \end{center}
  $^\dagger$: uncertain\\
  {\bf References.} (1) \citet{vanLeuven2007}, (2) \citet[][see the text for additional
  references]{Ducourant2014}, (3) \citet{Tokovinin2014}, (4) \citet{Tokovinin1999}, (5)
  \citet{Koerner2000}, (6) \citet{Boden2005}, (7) \citet{Soderblom1998}.
\end{table}

The presence of circumstellar material was revealed by the conspicuous infrared (IR)
excess in the system \citep{Walker1988, Zuckerman1993} and was later attributed to the B
component alone using high-resolution observations \citep[e.g.][]{Koerner2000,
  Prato2001}. The lack of emission at near-IR wavelengths, together with modeling of both
the Spectral Energy Distribution (SED) and architecture of the system, suggested the
existence of a small (a few au) cavity in the disk and a severely truncated outer radius
\citep[e.g.][]{Prato2001, Akeson2007, Pichardo2008}, as expected by the effect of the
Ba-Bb and A-B pairs, respectively \citep[e.g.][]{Artymowicz1994}. The dust composition was
first studied in detail by \citet{Furlan2007}, who presented the \emph{Spitzer}/Infrared Spectrograph
(IRS) of the source and found a broad silicate feature, indicating
the presence of large grains. They proposed that some optically thin dust is required
inside the disk cavity ($\approx 1.5-2\,$au) to account for the observed mid-IR spectrum,
and reproduced the continuum emission with an optically thick ring at
$\approx$6\,au. Later, \citet{Olofsson2012} suggested that a small fraction of crystalline
grains exist in the disk. The disk was finally (partially) resolved at 880\,$\mu$m by
\citet{Andrews2010a} using the Submillimeter Array (SMA) interferometer, finding a disk
extent of 10-15\,au. They did not detect CO(3-2) emission from the system.

The disk around HD~98800~B is peculiar; due to its age, lack of signatures of mass
accretion or molecular gas \citep{Soderblom1996, Kastner2004, Dent2005, Salyk2009}, and
deficit of near-IR excess, the object has usually been classified as an unusual debris
disk \citep{Verrier2008, Olofsson2012}. On the other hand, later detections of H$_2$
\citep{Yang2012} and [O\,I] \citep{Riviere-Marichalar2013}, together with its strong excess
in the far-IR with respect to typical the debris disk \citep[e.g.][]{Kennedy2012} also suggest
that the object could be in a transitional phase from protoplanetary to a debris disk
\citep[e.g.][]{Wyatt2007}. Given the typical lifetimes of protoplanetary disks around
single stars, the survival of a disk such as the one in HD~98800~B in a multiple system
(subject to strong dynamical interactions) has been challenging for disk evolution
theories.

%%%% OBSERVATIONS
\section{Observations}\label{sec:observations}

\subsection{VLA observations}

We present resolved Ka~band ($\approx 9$\,mm) and C~band ($\approx 5$\,cm) data of
HD~98800 obtained with the VLA of the National Radio Astronomy Observatory
(NRAO)\footnote{The NRAO is a facility of the National Science Foundation operated under
  cooperative agreement by Associated Universities, Inc.} within the Disks@EVLA program
(Project code: AC982, P.I.: Claire Chandler). Both sets of observations were performed with the
A~configuration, which provides the longest baselines (from 0.7 to 36.4\,km).

The Ka~band observations were carried out on 2012 October~27 and~28, with a total
on-source time of 169\,minutes. A total bandwidth of 2\,GHz was used, covering the
frequency ranges from 30 to 31\,GHz and from 37 to 38\,GHz. The C~band data were obtained
on 2011 July~6, with the correlator set up to cover the frequencies from
$\approx 4.3$\,GHz to $\approx 5.3$\, GHz and from $\approx6.8$ to $\sim7.8$ GHz,
resulting in a total bandwidth of 2\,GHz. The on-source time for the C~band was 20.5
minutes. At both bands, 3C286 was observed to perform the absolute flux calibration, while
J1112-2158 was used to obtain the gain calibration. Additionally, 3C279 was used as the
bandpass calibrator during the Ka~band observations.

\subsection{Spectral Energy Distribution (SED) compilation}\label{sec:SED_data}

To complement the VLA observations, we compiled photometry for both the A and B
components from several studies, including \citet{Rucinski1993}, \citet{Zuckerman1993},
\citet{Sylvester1996}, \citet{Soderblom1998}, \citet{Low1999}, \citet{Koerner2000},
\citet{Prato2001}, \citet{DiFrancesco2008}, and \citet{Andrews2010a}. We also included
data from the AKARI/IRC \citep{Ishihara2010} and AKARI/FIS catalogs. Most of the emission at
wavelengths $\geq 6 \mu$m arises from the disk around HD~98800~B \citep{Furlan2007};
therefore, we assigned all data at wavelengths longer than this value to the B component. The
photometry of HD~98800~B was then dereddened using the \citet{Mathis1990} extinction law
and $A_V=0.44$ \citep{Soderblom1998}. We also included the \emph{Spitzer}/IRS spectrum
(corrected from the contribution of the A component) presented in \citet{Furlan2007}.

\begin{deluxetable}{rrr}
\tablecaption{Photometry of HD~98800~A \label{tab:photometryA}}
\tablehead{
\colhead{Wavelength ($\mu$m)} & \colhead{Flux (mJy)} & \colhead{Reference}
}
\colnumbers
\startdata
0.429 & 266$\pm$8 & 1\\
0.525 & 700$\pm$20 & 1\\
0.954 & 2080$\pm$60 & 1\\
0.955 & 2000$\pm$60 & 1\\
1.082 & 2440$\pm$80 & 1\\
1.251 & 2000$\pm$200 & 2\\
1.452 & 2470$\pm$70 & 1\\
1.658 & 2300$\pm$200 & 2\\
1.875 & 2190$\pm$70 & 1\\
1.899 & 2300$\pm$70 & 1\\
2.20 & 1800$\pm$100 & 2\\
4.7 & 470$\pm$40 & 3\\
4.8 & 450$\pm$50 & 2\\
7.9 & 180$\pm$30 & 2\\
7.91 & 180$\pm$20 & 4\\
8.80 & 200$\pm$20 & 2\\
8.81 & 170$\pm$10 & 4\\
9.69 & 150$\pm$20 & 4\\
9.80 & 160$\pm$10 & 2\\
10.3 & 120$\pm$10 & 4\\
10.3 & 140$\pm$10 & 2\\
11.7 & 140$\pm$10 & 2\\
11.7 & 100$\pm$10 & 4\\
12.5 & 80$\pm$10 & 4\\
12.5 & 140$\pm$20 & 2\\
18.2 & 80$\pm$30 & 2\\
8820 & 0.080$\pm$0.015 & This work\\
49600 & 0.16$\pm$0.03 & This work\\
\enddata
\tablecomments{{\bf References.} (1) \citet{Soderblom1998}, (2) \citet{Prato2001}, (3) \citet{Zuckerman1993},
  (4) \citet{Koerner2000}.}
\end{deluxetable}

Additionally, the \emph{Herschel Space Observatory} \citep{Herschel} observed HD~98800
with both the Photoconductor Array Camera and Spectrometer (PACS) and the Spectral and
Photometric Imaging Receiver (SPIRE) instruments. For PACS, observations at 70\,$\mu$m
(obsid: 1342188473), 100, and 160 \,$\mu$m (obsids: 1342212634 and 1342212635) are
available (P.I.: B. Dent). We used the \emph{Herschel} Interactive Processing Environment
\citep[HIPE v.15,][]{Ott2010} to perform aperture photometry using standard values:
aperture radii of 12, 12, and 22\,arcsec for 70, 100, and 160\,$\mu$m, respectively;
background annulus radii of 35 and 45\,arcsec for all bands; and the corresponding
aperture corrections \citep{Balog2014}. For SPIRE, observations at 250, 350, and
500\,$\mu$m (obsid: 1342247269, P.I.: A. Roberge) were used to estimate photometric fluxes
using the recommended procedure of fitting the sources in the timeline
\citep{Pearson2014}. The \emph{Herschel} fluxes were attributed to HD~98800~B. We note
that the PACS fluxes listed in this study are 5-20\,\% higher than the ones reported in
\citet{Riviere-Marichalar2013}, which is likely a result of the updated HIPE version and
aperture correction factors used.

The photometry for the A and B components is presented in Tables~\ref{tab:photometryA}
and~\ref{tab:photometryB}, respectively. 

\begin{deluxetable}{rrr}
\tablecaption{Photometry of HD~98800~B \label{tab:photometryB}}
\tablehead{
\colhead{Wavelength ($\mu$m)} & \colhead{Flux (mJy)} & \colhead{Reference}
}
\colnumbers
\startdata
0.429 & 125$\pm$4 & 1\\
0.525 & 430$\pm$13& 1\\
0.954 & 1880$\pm$60 & 1\\
0.955 & 1890$\pm$60 & 1\\
1.082 & 2300$\pm$70 & 1\\
1.251 & 2200$\pm$200 & 2\\
1.658 & 2500$\pm$200 & 2\\
2.20 & 1790$\pm$100 & 2\\
4.68 & 720$\pm$40 & 3\\
4.80 & 580$\pm$60 & 2\\
7.90 & 430$\pm$50 & 2\\
7.91 & 440$\pm$20 & 2\\
8.80 & 890$\pm$50 & 2\\
8.81 & 830$\pm$30 & 4\\
9 & 1100$\pm$30 & 5\\
9.69 & 1620$\pm$70 & 4\\
9.80 & 1470$\pm$80 & 2\\
10.3 & 1900$\pm$80 & 4\\
10.3 & 1610$\pm$90 & 2\\
11.7 & 2030$\pm$110 & 2\\
11.7 & 2180$\pm$90 & 4\\
12.5 & 2020$\pm$80 & 4\\
12.5 & 2200$\pm$120 & 2\\
17.9 & 5000$\pm$300 & 4\\
18 & 6200$\pm$200 & 5\\
18.2 & 4900$\pm$500 & 2\\
20.8 & 5500$\pm$300 & 4\\
23.4 & 8500$\pm$400 & 6\\
24.5 & 8600$\pm$400 & 4\\
25 & 9700$\pm$800 & 7\\
60 & 7100$\pm$700 & 7\\
65 & 7000$\pm$2000 & \emph{AKARI}/FIS \\
70 & 7500$\pm$400 & This work \\
71 & 6300$\pm$1200 & 6 \\
90 & 5800$\pm$1300 & \emph{AKARI}/FIS \\
100 & 4300$\pm$300 & 7 \\
100 & 4600$\pm$200 & This work \\
156 & 2100$\pm$400 & 6\\
160 & 2500$\pm$100 &  This work\\
250 & 1120$\pm$40 &  This work\\
350 & 600$\pm$20 &  This work\\
450 & 500$\pm$200 & 8\\
500 & 310$\pm$12 &  This work\\
800 & 100$\pm$10 & 9\\
850 & 70$\pm$20 & 8\\
880 & 83$\pm$3 & 10\\
1100 & 63$\pm$6 & 11\\
1300 & 36$\pm$7 & 12\\
2000 & 25$\pm$3 & 11\\
8824 & 0.81$\pm$0.09 &  This work\\
49600 & 0.09$\pm$0.03 &  This work\\
\enddata
\tablecomments{{\bf References.} (1) \citet{Soderblom1998}, (2) \citet{Prato2001}, (3) \citet{Zuckerman1993}, (4) \citet{Koerner2000}, (5)
  \citet{Ishihara2010}, (6) \citet{Low2005}, (7) \citet{Low1999}, (8) \citet{DiFrancesco2008}, (9) \citet{Rucinski1993}, (10) \citet{Andrews2010a}, (11)
  \citet{Sylvester1996}, (12) \citet{Stern1994}.}
\end{deluxetable}

%%%% RESULTS
\section{Results}\label{sec:results}

\subsection{VLA Images}

We created synthesized VLA Ka~band (34\,GHz, 8.8\,mm) and C~band (6\,GHz, 5\,cm) images of
the HD~98800 system using a Briggs weighting (robust parameter=0.5), yielding 1$\sigma$
rms= 10 $\mu$Jy beam$^{-1}$, beam = $0.12$\,arcsec $\times$ 0.06\,arcsec and a beam with
a position angle (PA) = 79\,$^{\circ}$. A natural-weighted image (1$\sigma$ rms = 9\,$\mu$Jy beam$^{-1}$,
beam = $0.14$\,arcsec $\times$ 0.08\,arcsec, PA = 78\,$^{\circ}$) was also produced to
derive more reliable fluxes. The 8.8\,mm images clearly resolve the disk around HD~98800~B
and its inner cavity for the first time, revealing a disk structure suggestive of a
limb-brigthened ring with a diameter of $\approx$0.1\,arcsec. The A component is also
detected at 8.8\,mm as an unresolved source. At the C~band, only the Briggs-weighted image
(robust parameter = 0.5, 1\,$\sigma$ rms = 15\,$\mu$Jy beam$^{-1}$,
beam = 0.69\,arcsec $\times$ 0.28\,arcsec, PA = 23\,$^{\circ}$) separates both
components, but the disk is not resolved at this longer wavelength. Despite not having a
known disk around it, HD~98800~A is brighter than HD~98800~B at 5\,cm.

\begin{figure}
  \centering
  \includegraphics[width=8.1cm]{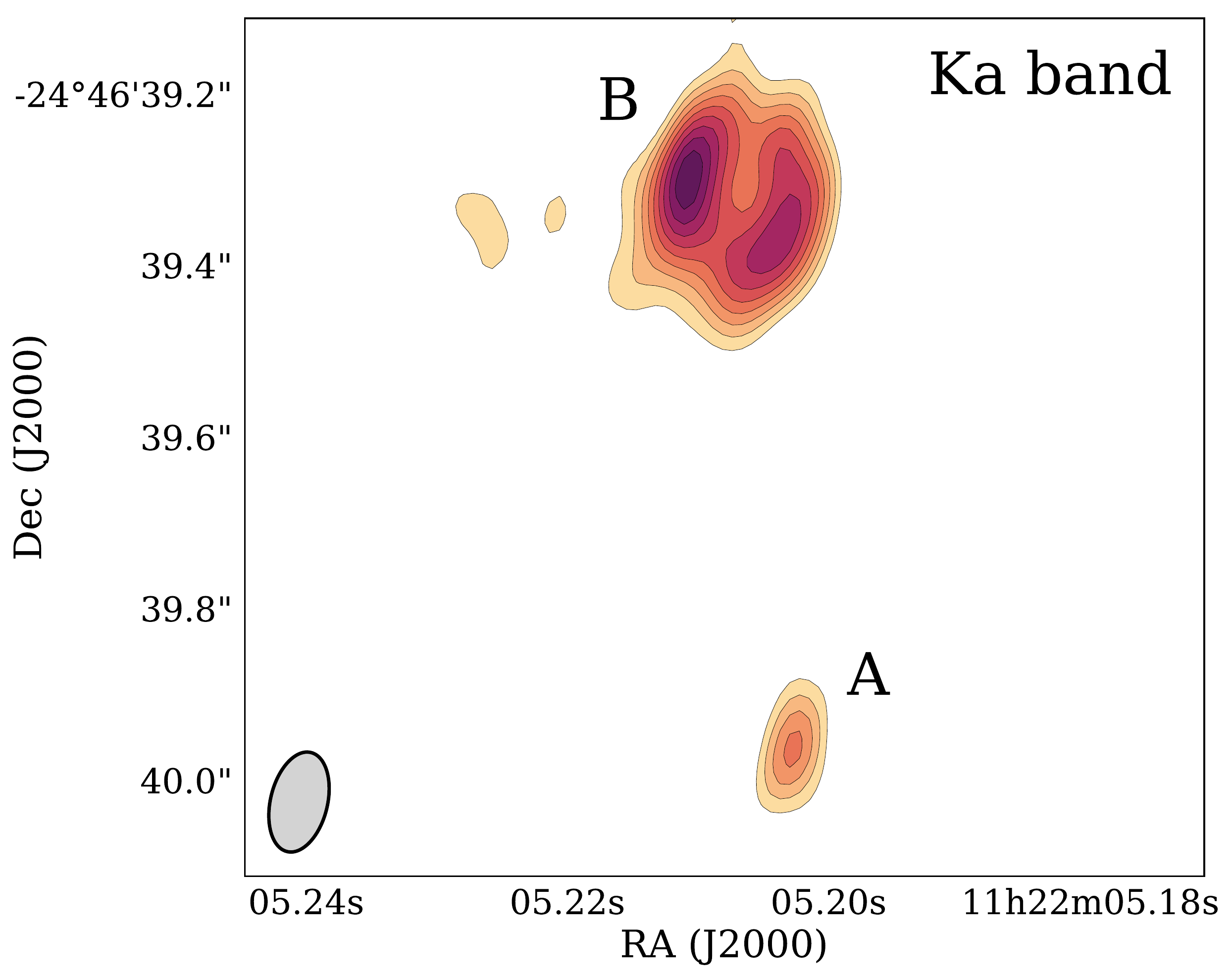}\\
  \includegraphics[width=7.8cm]{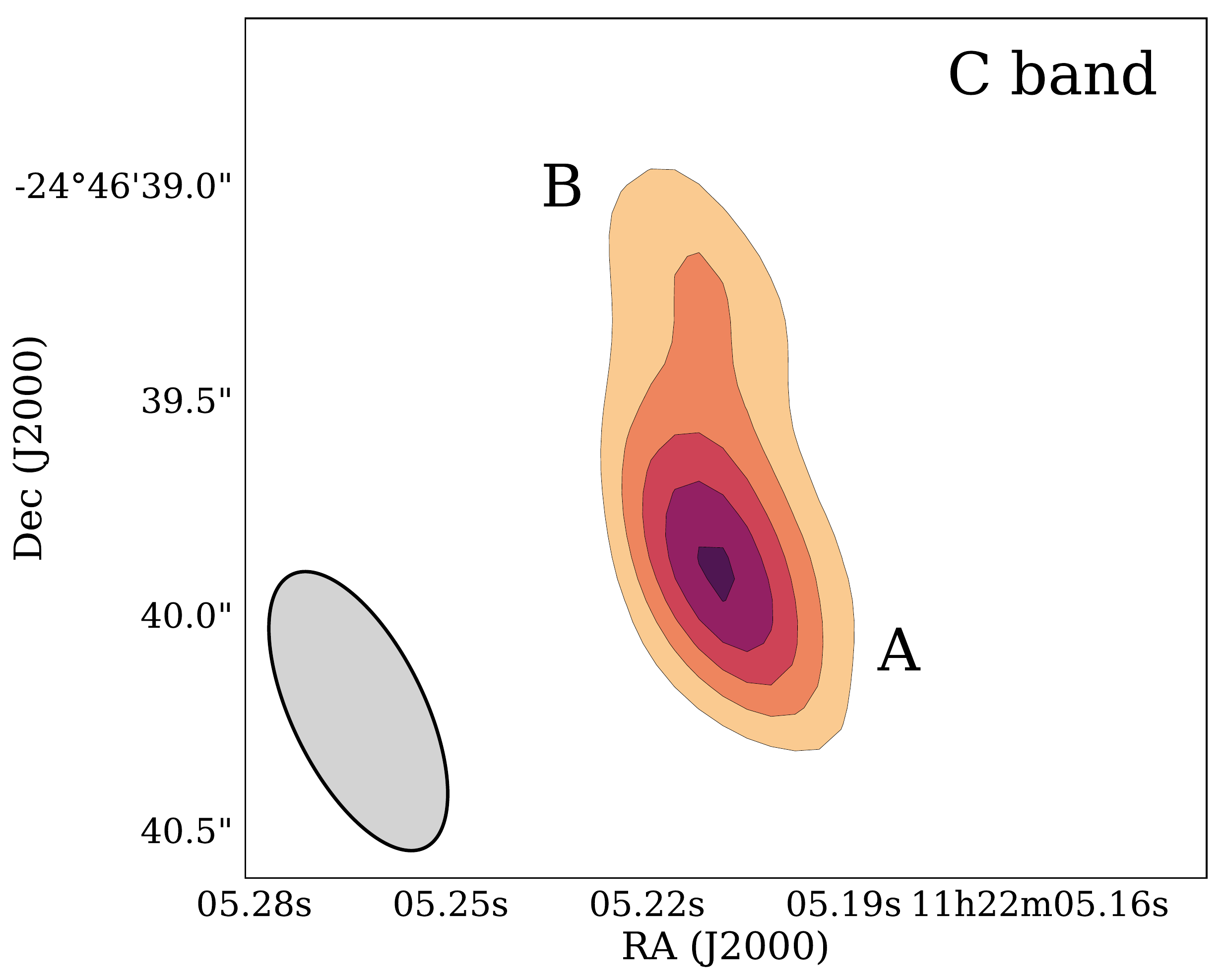}\\
  \includegraphics[width=7.8cm]{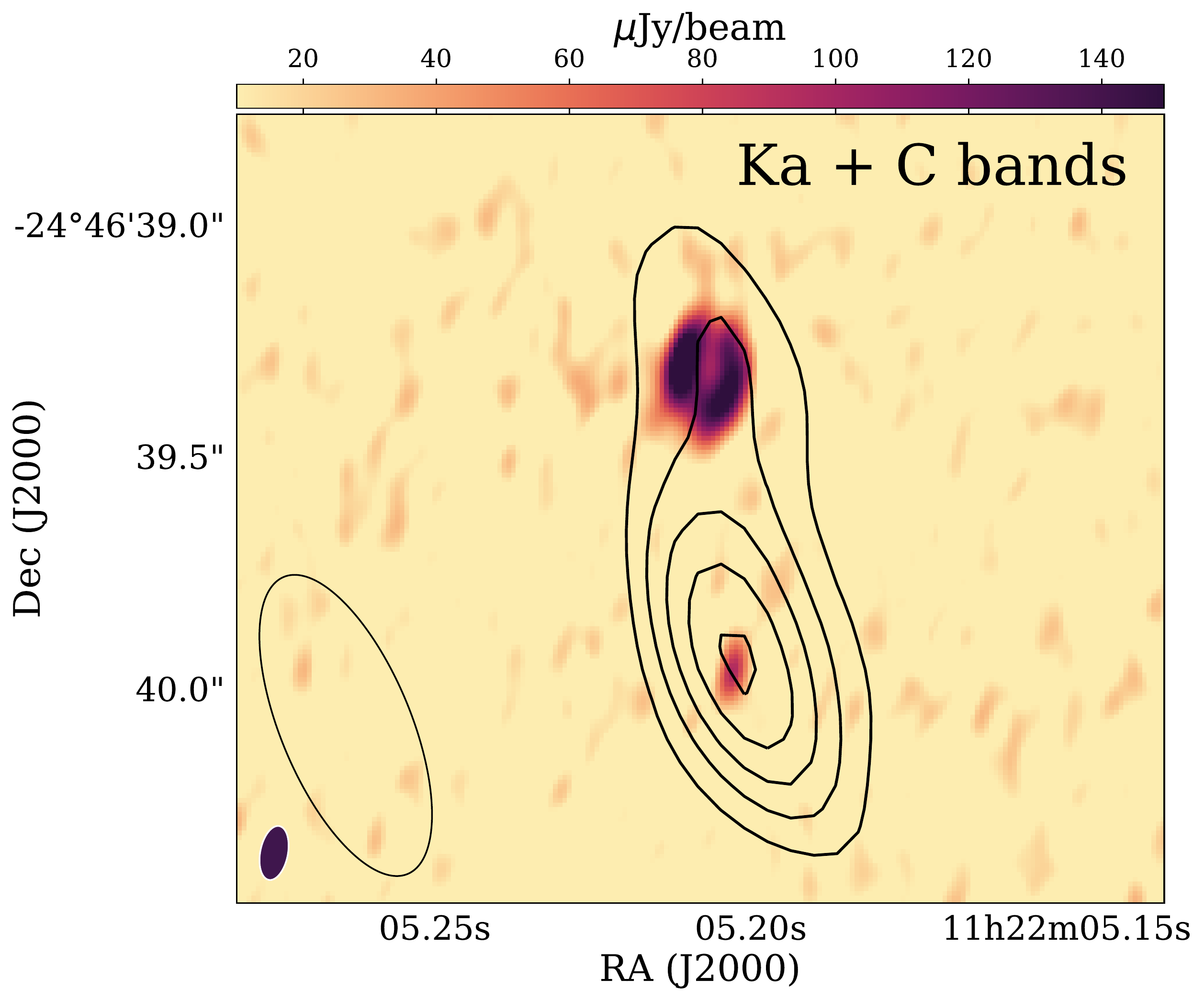}

  \caption{VLA observations of the HD~98800 system. Top: Ka~band (8.8\,mm) Briggs-weighted
    image (1$\sigma$ rms=\,10 $\mu$Jy beam$^{-1}$,
    beam=\,$0.12$\,arcsec$\times0.06$). Middle: C~band (5\,cm) Briggs-weighted image
    (1\,$\sigma$ rms=15\,$\mu$Jy beam$^{-1}$,
    beam\,=\,$0.69$\,arcsec$\times0.28$\,arcsec). In both cases, contours start at
    3\,$\sigma$ and increase in 2\,$\sigma$-steps. The corresponding beams are shown as
    gray ellipses. Bottom: Ka~band image with overlaid contours from the C band, for
    comparison. The C band images have been corrected from proper motion. The
    corresponding beams are shown in the bottom left corner (filled ellipse for the
    Ka~band, empty ellipse for the C~band).}\label{fig:VLA_maps}
\end{figure} 

Figure~\ref{fig:VLA_maps} shows the Briggs-weighted Ka~band and C~band images, and an
overlay of both. For comparison, we have applied a correction to the C~band images to
account for the measured proper motion of the HD~98800 system
\citep[$\mu_\alpha cos(\delta)$=85.4 mas\,yr$^{-1}$, $\mu_\delta$=33.1
mas\,yr$^{-1}$,][]{vanLeuven2007}.

\newpage
\subsection{VLA fluxes}
 
We used the natural-weighted image at Ka~band to derive fluxes both for HD~98800~A and B,
since the components are well separated. Their fluxes were estimated by repeatedly
measuring them in different regions placed around each source. This yielded fluxes of
70$\pm$10$\mu$Jy for~A and of 810$\pm$90\,$\mu$Jy for B. Since the A component is
unresolved, we also fitted a 2D Gaussian to it using the Common Astronomy Software
Applications package \citep[CASA,][]{CASA}, and derived a total integrated flux of
80$\pm$10\,$\mu$Jy and a peak flux of 90$\pm$10\,$\mu$Jy (the resulting fit confirmed that
A is unresolved). We adopted a final value for the flux of HD~98800~A of
80$\pm$15\,$\mu$Jy at 34\,GHz.

At the C~band, natural weighting barely separates the two components, and thus we used
Briggs weighting to estimate their 5\,cm fluxes. In this case, the flux of each source was
calculated by fitting two Gaussians to the observed emission using the task \emph{imfit}
in CASA, yielding integrated and peak fluxes of 160$\pm$30\,$\mu$Jy and
170$\pm$20\,$\mu$Jy for A, and 100$\pm$30\,$\mu$Jy and 75$\pm$15\,$\mu$Jy for B,
respectively. Therefore, we assigned a final 6\,GHz flux of 160$\pm$30\,$\mu$Jy to
HD~98800~A and 90$\pm$30\,$\mu$Jy to HD~98800~B.

Circular polarization was also searched for using the VLA observations. We processed the
right and left polarization images both at Ka and C~bands using the Briggs-weighted
images. At the Ka~band, the polarization fluxes were determined to be 100$\pm$25\,$\mu$Jy
(right) and 90$\pm$25\,$\mu$Jy (left) for A and 790$\pm$50\,$\mu$Jy (right) and
770$\pm$70\,$\mu$Jy (left) for B. In both polarizations, the disk around HD~98800~B is
resolved, with no hint of emission from a central source. At C band, the corresponding
fluxes are 200$\pm$30\,$\mu$Jy (right) and 140$\pm$20\,$\mu$Jy (left) for A, and
80$\pm$20,$\mu$Jy (right) and 70$\pm$20\,$\mu$Jy for B. Thus, only the 5\,cm emission of
HD~98800~A shows signatures of some polarization in the VLA data, which could indicate the
presence of nonthermal gyrosynchroton emission.

In all cases, the reported uncertainties include the 10\,\% systematic VLA flux
calibration uncertainty.

\begin{figure}
  \centering
  \includegraphics[width=\hsize]{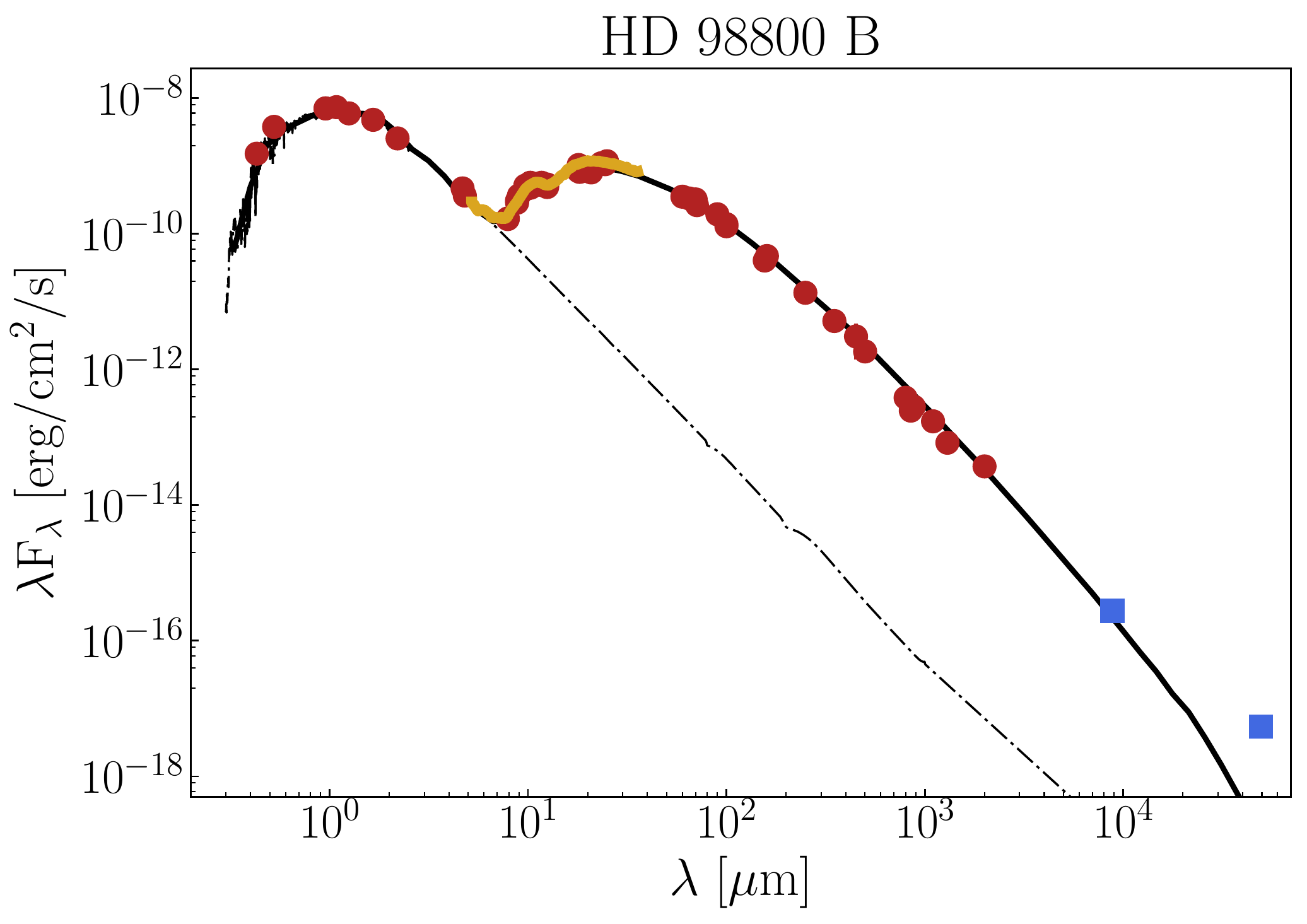}
  \includegraphics[width=\hsize]{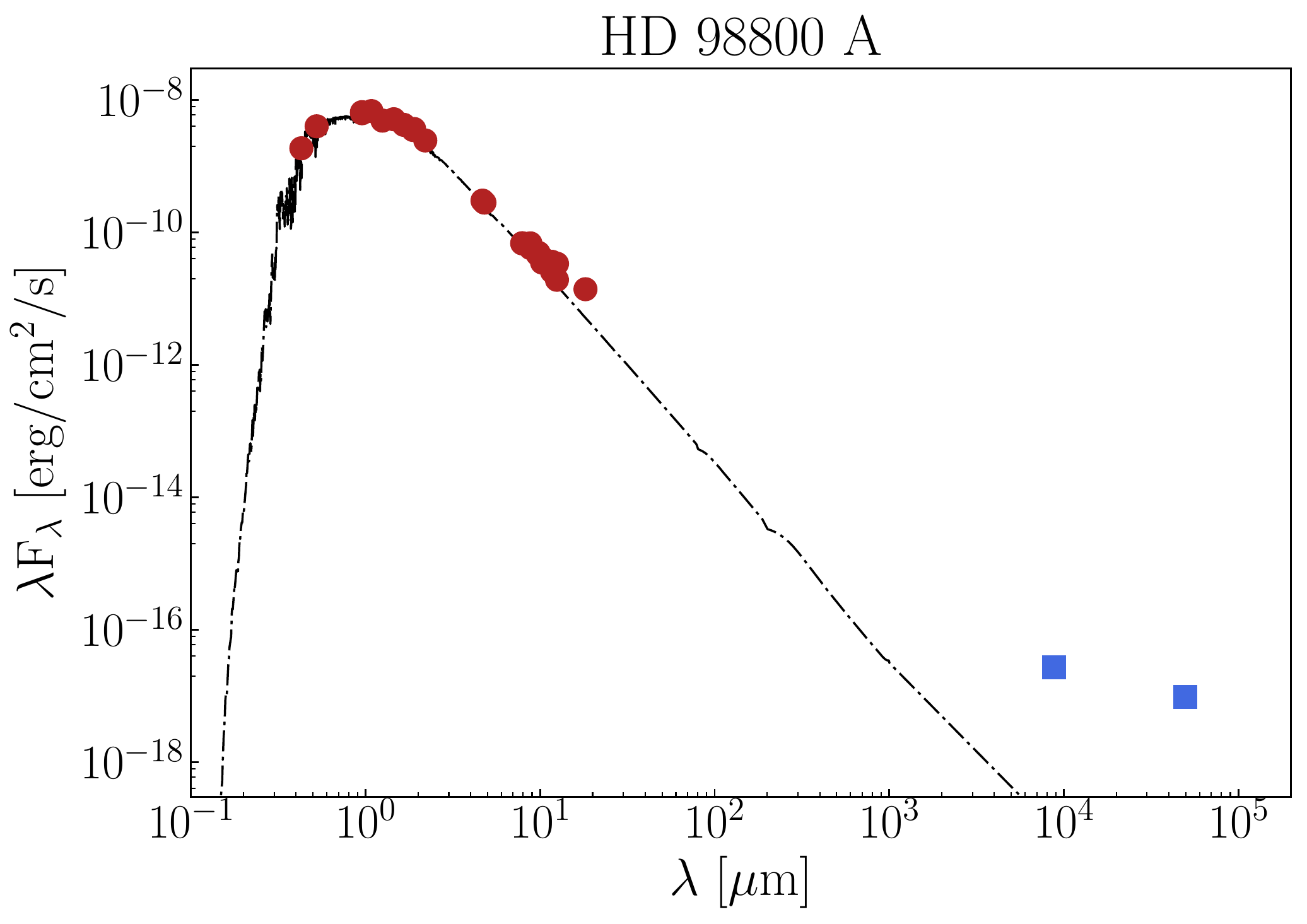}
  \caption{SEDs of HD~98800~B (top) and HD~98800~A (bottom). Ancillary (red
    dots) and VLA (blue squares) photometry are shown. Uncertainties are smaller than
    the symbol sizes. The \emph{Spitzer}/IRS spectrum of HD~98800~B (orange line) is also
    shown. Dashed lines are the corresponding stellar photospheres. In the case of
    HD~98800~B, the best-fitting disk model is plotted as a solid black line. The adopted
    model reproduces the shape of the SED (in particular, its spectral index) up to the
    VLA Ka-band observations (8.8\,mm). The C-band (5\,cm) VLA flux is significantly above
    the photospheric emission and is too shallow to originate from dust thermal emission,
    suggesting that an additional emission mechanism becomes important at longer
    wavelength. In the case of A, the tentative excess at 18\,$\mu$m like originates from
    uncertainties in separating the contribution of each components in unresolved
    photometry \citep[see][]{Prato2001}.}\label{fig:SEDs}
\end{figure}

The complete SEDs of HD 98800~A and B, including the new VLA fluxes, are shown in Figure~\ref{fig:SEDs}.

\subsection{Spectral indices}\label{sec:spectral_indices}

The millimeter spectral index $\alpha$ of the disk emission ($F_\nu \propto \nu^\alpha$)
is informative of some of its properties. If the emission is optically thin, $\alpha$
traces the maximum size of dust grains, with $\alpha>3$ indicating small sizes (e.g.,
interstellar medium (ISM)-like grains have $\alpha=3.6-3.8$), while values closer to
$\alpha=2$ suggest grain growth to mm/cm sizes \citep{Natta2004a, Ricci2010_Ophiuchus,
  Ricci2010_Taurus, Ribas2017}. On the other hand, optically thick emission will also
display $\alpha=2$, corresponding to blackbody radiation.

\begin{figure}
  \centering
  \includegraphics[width=\hsize]{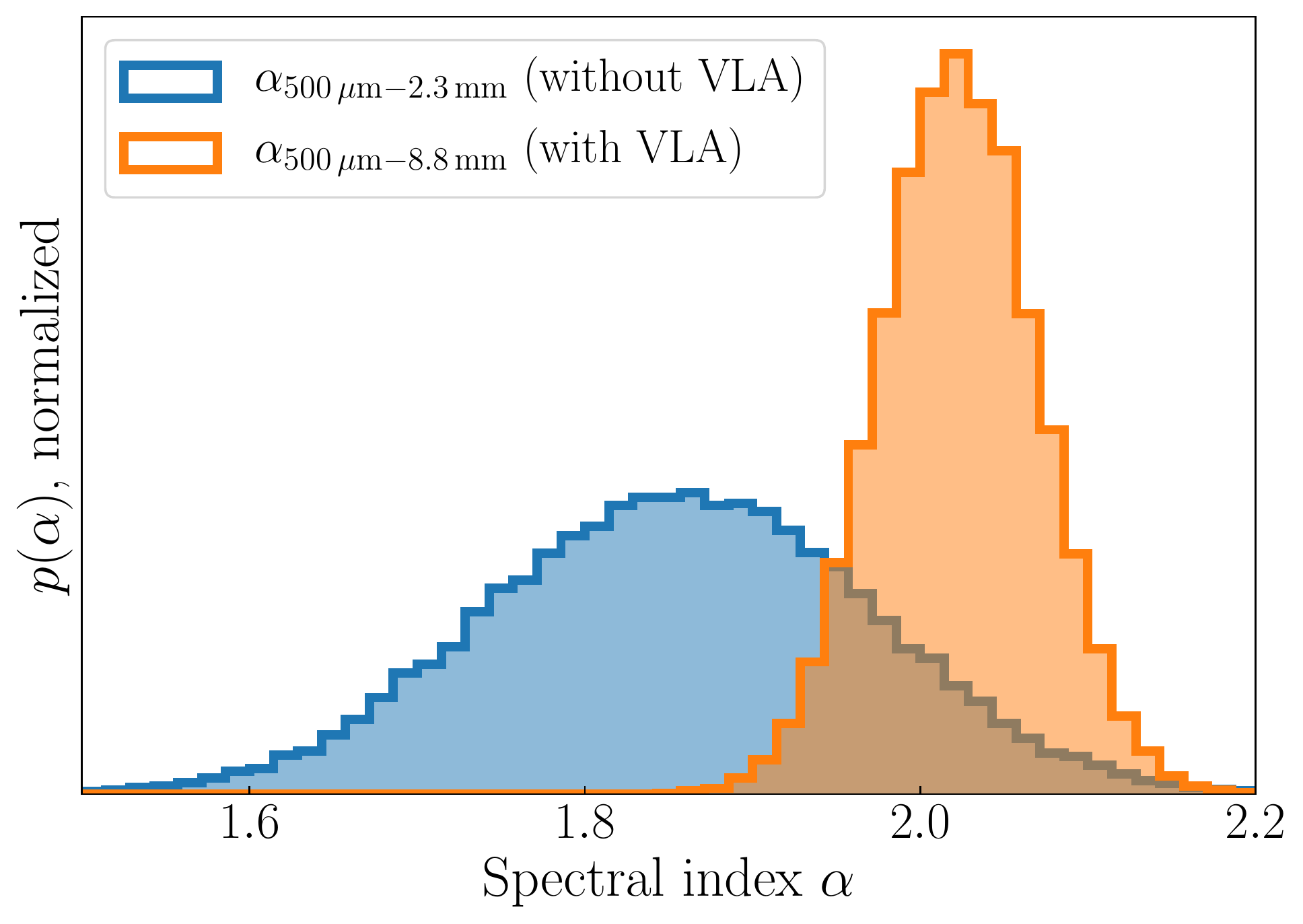}
  \caption{Posterior distributions for the millimeter spectral index $\alpha$, derived between
    500\,$\mu$m and 2.3\,mm (without VLA data, blue) and between 500\,$\mu$m and 8.8\,mm
    (with VLA data, orange).}\label{fig:spectral_indices}
\end{figure}

We estimated the millimeter spectral index $\alpha$ of the disk around HD~98800~B
by fitting a power law to the observed fluxes in the logarithmic scale
  ($\log{F_\nu} = \alpha \log{\nu} + C$) using the implementation of the Affine
Invariant Markov Chain Monte Carlo (MCMC) \citep{Goodman2010} in the \emph{emcee}
\citep{emcee} Python package. Uniform priors where adopted for both parameters
  ($\alpha$ and $C$). The minimum photometric uncertainty was set to 10\,\% to avoid
excessive weighting of individual data. The spectral index between 500\,$\mu$m and 2.3\,mm
(without including the new VLA data) is $\alpha = 1.9_{-0.1}^{+0.2}$ and becomes
$\alpha=2.02_{-0.05}^{+0.1}$ when the VLA flux at 8.8\,mm is included, which is consistent with
either optically thick emission or optically thin emission from large grains (quoted
  values and uncertainties correspond to the median and the 16- and 84-percentiles,
  respectively). The obtained posterior distributions are shown in
  Figure~\ref{fig:spectral_indices}.  The results do not change significantly if a
different wavelength range is used (e.g., $\alpha=2.09_{-0.07}^{+0.14}$ from 1 to
8.8\,mm).

We did not include the C~band VLA flux in the estimate of the spectral index of
HD\,98800~B; for wavelengths longer than 5-7\,mm, additional contribution from mechanisms
other than thermal dust emission, such as free-free emission from photoevaporative disk
winds or gyrosynchrotron radiation from stellar activity, can become significant
\citep{Pascucci2012, MacGregor2015, Macias2016, Booth2017}. Indeed, the spectral index
between the VLA Ka and C bands is $\alpha=1.3_{-0.2}^{+0.4}$, showing that there is
contribution from (at least) one of these processes. In the case of a photoevaporative
wind, unpolarized emission with a spectral index of $-0.1>\alpha>0.6$ is expected, whereas
gyrosynchrotron radiation is highly variable, can appear polarized, and has a negative
$\alpha$ value. Due to the lack of observations between 8.8\,mm and 5\,cm, it is not
possible to determine the wavelength at which the break in the spectral index occurs, and
thus we cannot distinguish between these explanations. However, given the nondetection of
polarization, the presence of gas in the system and the proposed evolutionary scenario,
it is likely that photoevaporation is taking place in the disk. In that case, it is also
possible that the 8.8\,mm flux has some contribution from free-free emission. If this
contribution is important, the resolved VLA images may be partially tracing the
photoevaporative wind, making HD~98800~B a great source for studying the photoevaporation
of protoplanetary disks. Follow-up observations that improve the wavelength coverage at
millimeter wavelengths will be able to quantify the relevance (if any) of free-free
emission at 8.8\,mm and test this idea in further detail.

In contrast with HD~98800~B, the A component shows no infrared excess and it is thought
to be diskless \citep[e.g.][]{Koerner2000, Prato2001}. However, both the Ka~and C~band VLA
fluxes are clearly above the expected photospheric level and the spectral index between
these two bands is $\alpha=-0.40_{-0.15}^{+0.30}$, which, together with the partial
polarization measured at 5\,cm, suggests that either free-free emission from a previously
unknown disk or chromospheric activity (or both) are present in HD~98800~A.

\subsection{A disk model of HD~98800~B}\label{sec:model}

While general estimates of the disk structure can be obtain directly from the resolved
images, the elongation of the beam makes it difficult to draw more precise values for some
morphological parameters. We therefore used a disk model to estimate some of its
properties by simulating the VLA observations. Our aim was to determine the disk spatial
parameters (i.e., inner and outer radii, inclination, and PA) while
reproducing the overall shape of the SED. We did not attempt to fit the dust mineralogy;
for studies of the dust composition in HD~98800~B, see \citet{Furlan2007} and \citet{Olofsson2012}.

\begin{figure*}
  \centering
  \includegraphics[height=2.13in]{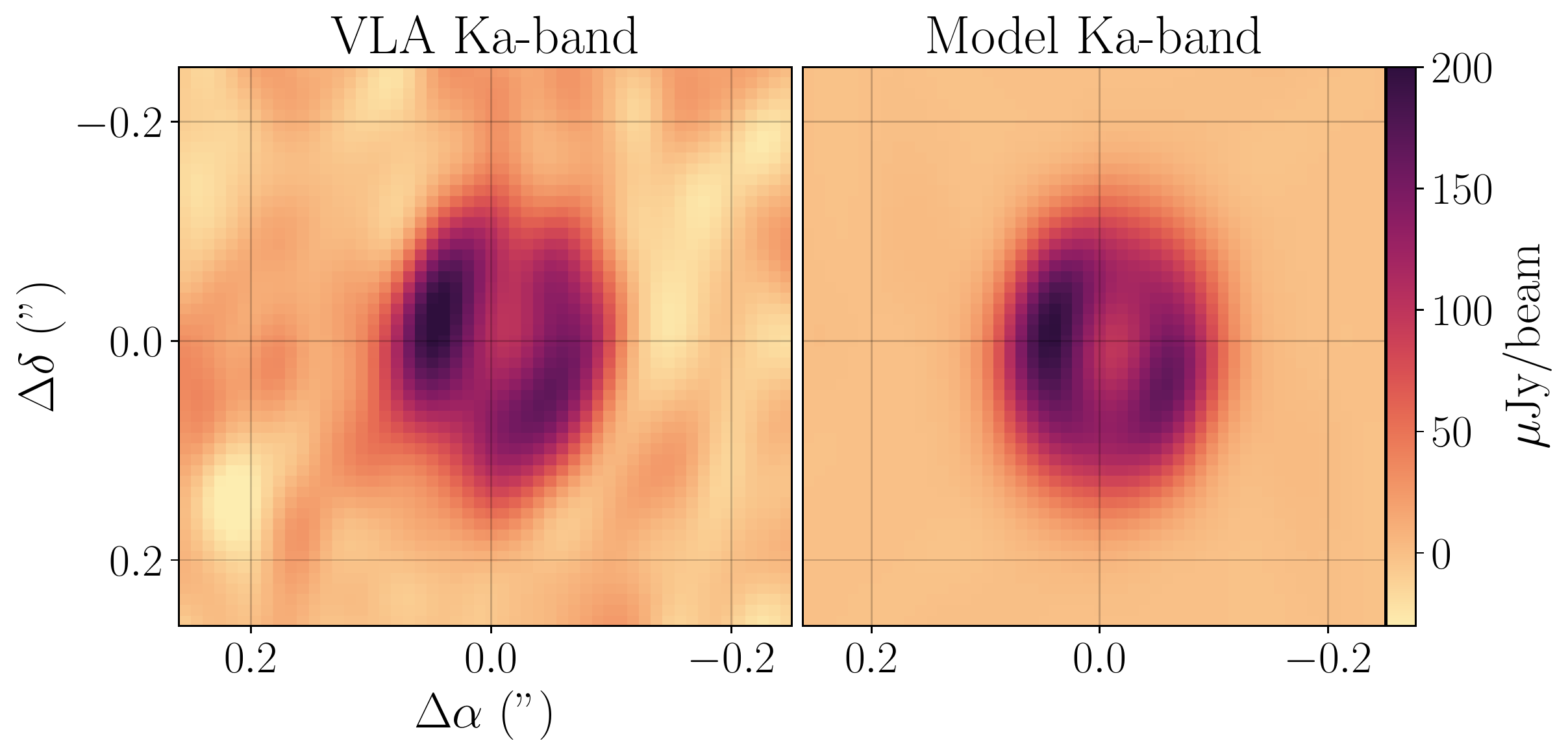}\hfill\includegraphics[height=2.13in]{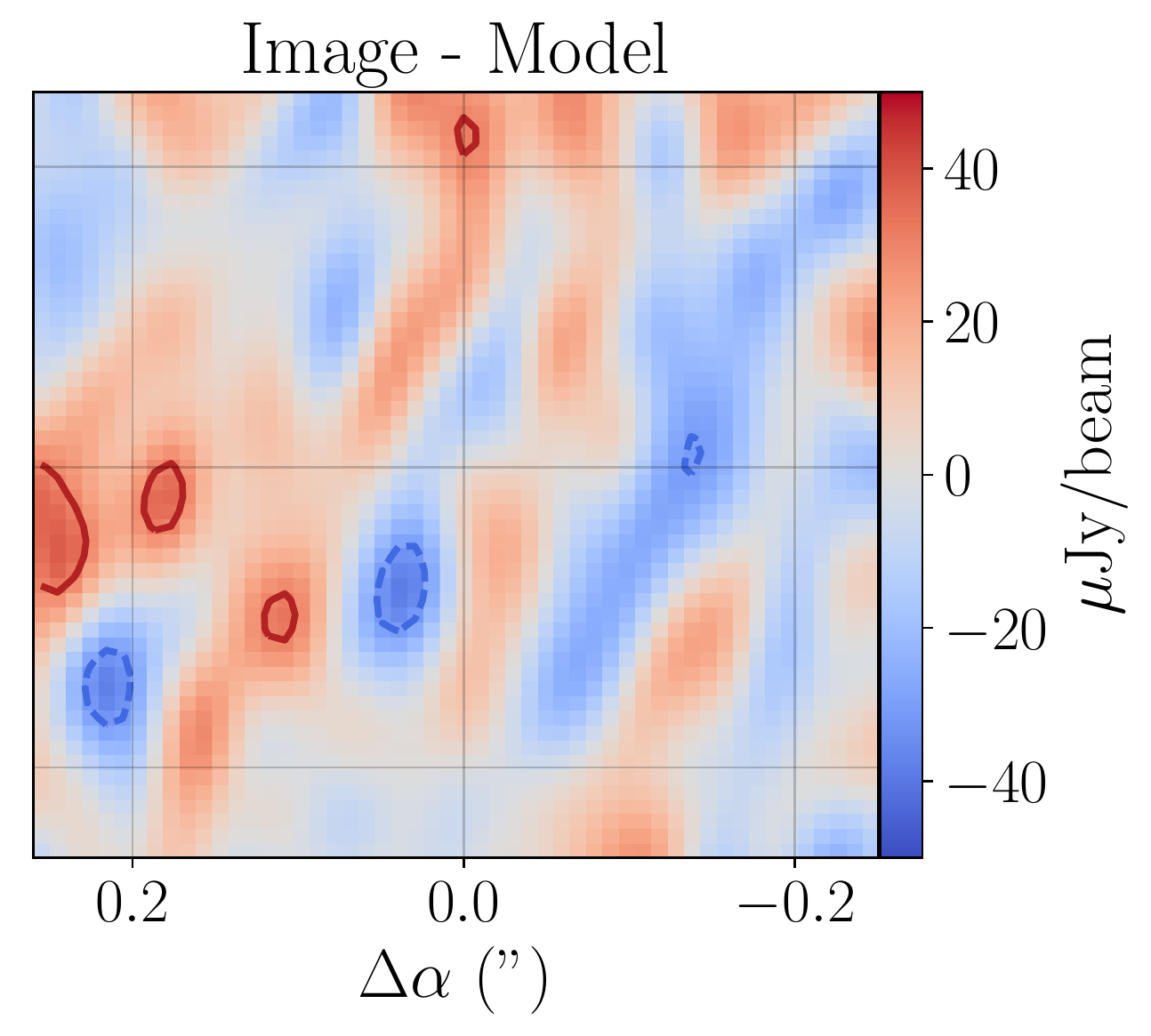}
  \caption{VLA Ka-band observation (left) of the disk around HD~98800~B , and
    best-fit simulated image (middle). The color scale ranges from -3$\sigma$ to
    20$\sigma$. Right: residuals of the best-fit model. Positive values are
    shown in red, negative are shown in blue. Dashed blue and red solid contours show structures
    below and above $-$3$\sigma$ and 3$\sigma$, respectively. In this case, the color scale ranges
    from -5$\sigma$ to 5$\sigma$.}\label{fig:model}
\end{figure*}

We used the radiative transfer code MCFOST \citep{MCFOST} to produce the disk model. The
distance was assumed to be 45\,pc \citep{vanLeuven2007, Donaldson2016}. We set up two
stars to account for the two components of HD~98800~B; component Ba with $T_*$=4200\,K,
$R_*$ = 1.1\,$R_\odot$, $M_*$ = 0.7 $M_\odot$; and Bb with $T_*$=4000\,K, $R_*$ =
0.9$\,R_\odot$, $M_*$ = 0.6 $M_\odot$, separated by a distance of 1\,au
\citep[][]{Boden2005}. As a starting point, we made crude estimates of some disk
parameters from the image, setting the disk inner and outer radii to 3 and 5\,au,
respectively, the inclination to $\approx$45\,$\deg$, and the PA to
$\approx$0\,$\deg$. The dust grain size distribution was assumed to follow
$n(a)\propto a^{-p}da$, where $a$ is the grain size, and we set $p=3$ based on the results
of \citet{Olofsson2012}. We adopted an astronomical silicates composition for the dust
\citep{Draine1984}. The observed spectral index $\alpha=2$ required a significant amount
of mass of large particles, and we obtained a good overall SED fit with a dust mass of
5$\times 10^{-5}$\,$M_\odot$ and minimum and maximum dust grain sizes from 1\,$\mu$m to
1\,cm. The minimum grain size in protoplanetary disks is usually assumed to be much
smaller, $a\approx0.005\,\mu$m, as in the ISM. However, given the age of the system,
significant dust growth is expected in the disk and is in agreement with the aforementioned
broad silicate feature. Expanding the grain size distribution to smaller sizes does not
affect the shape of the millimeter emission significantly, but much more mass is required
to account for the observed flux levels. The surface density profile used follows
$\Sigma(r)\propto r^{-1}$. We then changed the scale height of the disk so that it
  matched the shape of the observed SED, affecting mostly the mid- and far-IR emission:
  adopting a flaring law $h(r)=h_0 (r/10\,au)^{1.1}$, $h_0$=0.6\,au at 10\,au produced the
  best results. We note that this value probably does not represent the vertical structure
  of the disk accurately because our models do not include dust settling
  \citep[e.g.,][]{Dullemond2004_settling}, which concentrates large grains in the
  midplane. Given the additional complexity of this process, we chose not to incorporate
  it in our modeling, because it will not impact the results for the spatial parameters of
  the disk (see below).  However, a more settled midplane would also be optically thicker,
  and thus a settled disk with a lower mass could also produce the observed spectral index
  of $\approx 2$. For completeness, we also included a small (2.5-2.6\,au), optically
thin ring with 8$\times$10$^{-3}$\,$M_{\rm Moon}$ of small dust (0.25-1\,$\mu$m) to
reproduce the overall shape of the IRS spectrum, although this component is not relevant
for the mm-cm emission of the system.

Given the complex structure of the multiple system, we do not aim at fitting the VLA
visibilities. Instead, we used MCFOST and the parameters used in the SED model to produce
a grid of 625 disk images by varying the inner and outer radii ($R_{\rm in}$ and $R_{\rm out}$),
inclination ($i$), and PA of the disk. Five different values were tried
for each parameter: $R_{\rm in}$ from 2.5\,au to 3.5\,au in steps of 0.25\,au, $R_{\rm out}$
from 5\,au to 6,\,au in steps of 0.25\,au, $i$ from 30$^\circ$ to 50$^\circ$ in steps of
5$^\circ$, and PA from -20$^\circ$ to 20$^\circ$ in steps of 10$^\circ$. These ranges
were chosen based on visual inspection of the VLA image. We Fourier transformed the
synthetic model images using the same uv coverage as the observations to obtain the
simulated visibilities. Then, we obtained cleaned model images using the CLEAN task in
CASA. Since our SED models do not exactly reproduce the VLA flux at 8.8\,mm, we rescaled
the simulated images to the measured flux in the VLA 8.8\,mm Briggs-weighted image. The
residuals for each image were computed, and the resulting best fit is shown in
Figure~\ref{fig:model}. We used the best 10 models to derive parameter ranges from our fitting procedure.  This
process yielded $R_{\rm in}$ = 2.75-3.25\,au \citep[in agreement with the SED modeling results
of][]{Akeson2007,Andrews2010a}, $R_{\rm out}$ = 5-5.5\,au, $i$ = 40$^\circ$-45$^\circ$, and
PA = $0^\circ-10^\circ$. We note that there is a small -3\,$\sigma$ residual in the southeast
part of the disk. Although this might indicate a deficit of emission at that position in the
disk, the spatial resolution and sensitivity of the observations do not allow us to draw
any conclusion on its origin, and we do not discuss it here. Future high
spatial-resolution observations will probe finer details of the disk structure and reveal
if the structure is real.

The derived disk outer radius ($5-5.5$\,au) is smaller than the disk size of 10-15\,au
obtained by \citet{Andrews2010a} using SMA observations at 880\,$\mu$m. Large dust grains
in protoplanetary disks are subject to radial migration due to the drag force exerted by
the gas \citep[][]{Adachi1976, Weidenschilling1977, Takeuchi2002}, a process that
concentrates large grains in the inner regions of the system and results in
wavelength-dependent disk sizes. This effect has already been observed in several cases
\citep[e.g.][]{Guilloteau2011,Perez2015,Tazzari2016}, with disks appearing smaller at
longer wavelengths. While it is possible that radial migration is the origin of the
different disk-size estimates at 880\,$\mu$m (SMA) and 8.8\,mm (VLA) observations, we note
that the SMA resolution was significantly lower (with a beam size of 0.92\,arcsec $\times$
0.68\,arcsec) and the disk was only marginally resolved. Additionally, the PA from SMA
observations ($160^\circ$) is also different than the one suggested by the VLA
observations ($0^\circ-10^\circ$). We thus consider that the most probable explanation for
the difference in disk size is that the resolution of the SMA observations was not
sufficient to derive an accurate disk radius, although radial migration cannot be ruled
out. Observations at (sub)mm wavelengths with higher angular resolution will provide more
accurate disk-size estimates at shorter wavelengths and will confirm dust radial migration in
HD~98800~B, if present.

%%%% DISCUSSION
\section{Discussion}\label{sec:discussion}

\subsection{HD~98800~B: likely a protoplanetary disk}

  The nature of HD~98800~B is uncertain given its properties and the current understanding
  of disk evolution. It has usually been considered as a young, peculiar debris disk
  \citep[e.g.][]{Verrier2008,Olofsson2012} based on its age
  \citep[$\approx$10\,Myr,][]{Webb1999, Weinberger2013} and the fact that is well fit with
  a single blackbody \citep[e.g.][]{Furlan2007, Kennedy2013}. However, some of its
  properties do not match those of debris disks, which has typically been interpreted as a
  result of HD~98800~B being an intermediate stage between a protoplanetary and a debris
  disk. The presented VLA observations provide additional information about the source:

\begin{enumerate}

\item The fractional luminosity, $f$, of a disk is defined as the total luminosity
  of the disk with respect to that the central object. For HD~98800~B, $f$ is found to be
  $\approx$20\,\% based on our SED model, which is in agreement with previous works
  \citep[10-20\,\%,][]{Furlan2007, Kennedy2013}. In contrast, debris disks display much
  lower fractional luminosities, typically $f\lesssim1\,\%$ \citep[e.g.][]{Wyatt2008,
    Matthews2014, Hughes2018}.

\item A fractional luminosity of 20\,\% requires that the solid angle of the disk, as seen
  from the star, is large. This needs a large opening angle/flaring which, in turns,
  favors the idea that the structure of the disk is dominated by gas. While an increasing
  number of debris disks have been found to harbor gas \citep[e.g.][]{Moor2017,
    Pericaud2017}, their content is much lower than those of protoplanetary ones.

\item Such a high fractional luminosity also requires that the dust is mostly optically
  thick at optical/near-IR wavelengths, where the bulk of stellar radiation is emitted --
  otherwise, even with a large solid angle, the disk emission would be weaker. In contrast,
  debris disks are optically thin at all wavelengths \citep{Hughes2018}.

\item The detection of [O\,I] \citep{Riviere-Marichalar2013, Riviere-Marichalar2016} and
  H$_2$ \citep{Yang2012} directly confirms the presence of gas content in the disk. In
  particular, H$_2$ has not been found in any debris disk so far because it is not expected
  to arise from planetesimal collisions and will, therefore, trace primordial gas.

\item The spectral index of HD~98800~B is $\alpha=2$ up to 8.8\,mm. Such a spectral index
  could indicate either optically thick emission or an optically thin disk with a
  gray opacity law in the millimeter. However, most debris disks show spectral indices of
  $\alpha=2.5-3$ \citep{Holland2017, Sibthorpe2018}, which are inconsistent with the one observed in
  this source. Together with the other features, this suggests that the disk emission is,
  in fact, optically thick at millimeter wavelengths.

\end{enumerate}

The combination of these pieces of evidence suggests that the disk around HD~98800~B is
probably not a debris disk, but instead may be a massive (optically thick) and gas-rich
\emph{ring}; a result of the gravitational interactions of both the Ba+Ba pair and the A
companion. Using the typical gas-to-dust mass ratio of 100 in protoplanetary disks, our
combined SED+image modeling yields a disk mass of 5\,$M_{\rm Jup}$ (albeit largely
uncertain, because the dust mass could be both higher due to the emission being optically
thick and lower if settling is included in the modeling). This disk would have a high
density, and planet formation may thus still be possible in the system. Additionally, such
a compact, optically thick ring would also explain the lack of CO detection by previous
studies \citep{Zuckerman1995, Andrews2010a}, given its small surface area. Follow-up
observations of this target with the Atacama Large Millimeter/submillimeter Array (ALMA)
will search for CO in the disk with higher sensitivity levels, testing this hypothesis.

\subsection{Implications for disk evolution in HD~98800}

Given the 10\,Myr age of HD~98800~B, it is surprising that the system still retains a
massive disk.  Most protoplanetary disks around single stars disperse within that
timescale, but binary and multiple systems have complex dynamical interactions than can
disperse their disks much faster \citep[e.g.][]{Cieza2009, Harris2012, Kraus2012}. Recently,
\citet{Rosotti2018} have shown that the presence of a close companion ($a<$20-30\,au) can
even alter the normal evolution of the disk from the standard inside-out dispersal (as in
single stars) to an outside-in evolution, also resulting in reduced lifetimes. However,
multiplicity may also be the key to explain the long-lived disk around HD~98800~B.

A number of papers have studied the dynamics of the system as a debris
disk. \citet{Akeson2007} modeled it as a triple system (Ba-Bb-A) by distributing test
particles from 2 to 15\,au around the B pair, and the resulting disk extended from 3 to
10\,au (and also showed substantial warping due to the effect of A). The modeling of
\citet{Verrier2008} suggested three different structures (a prograde coplanar disk, a
second retrograde disk, and a surrounding halo) extending from 4 to 15\,au. Interestingly,
the VLA images show a quite symmetric, featureless disk, although future observations with
higher resolution may reveal smaller structures. Later, numerical simulations by
\citet{Domingos2012} predicted outer disk radii ranging from 7.8 to 12.2\,au, depending on the
orbital parameters used. While of great interest for the study of particle dynamics in
multiple systems, we note that all of these works derived outer radii larger than the value
obtained from the resolved VLA 8.8\,mm image. A feasible explanation for this is that none
of these studies included gas in their simulations, leading to a different disk behavior
and evolution.

Companions can play a major role in the evolution of a gas-rich disk through tidal forces,
both from the circumprimary (a disk surrounding one of the components) and circumbinary
(a disk surrounding both stars) points of view.

\begin{enumerate}

\item Binaries surrounded by a circumbinary disk transfer angular momentum to the material
  in the inner regions of the disk through tidal torques, accelerating this material and stopping its
  inward motion; this process truncates the inner disk and opens up a gap. As an
  example, \citet{Artymowicz1994} found that, for a reduced mass of $\mu$=0.3, the disk is
  truncated at 2$a$-2.5$a$ (where $a$ is the semi-major axis of the binary) for
  eccentricities $e=$0 and 0.4, respectively. Considering that the truncation radius
  increases for higher eccentricities and the orbital parameters of HD~98800 B
  \citep[$a_{Ba-Bb}=0.5$\,au, $\mu=0.45$ and $e\sim0.8$,][]{Boden2005}, a cavity of
  $\sim$3-3.5\,au is not surprising for this system. \citet{Pichardo2008} modeled
  HD~98800~B using the same orbital parameters while including the effect of gas and
  predicted a $\sim$3\,au gap in the disk that is in total agreement with
  the presented observations. The effect of the inner binary also reduces the accretion
  rate onto the central sources drastically (or completely), resulting in longer-lived
  disks around close binaries with respect to single stars \citep{Alexander2012}.

\item On the other hand, the torque exerted by the external companion of a circumprimary
  disk removes angular momentum from its outer regions, hence slowing (or completely
  stopping) its viscous diffusion; circumprimary disks have truncated outer
  radii. \citet{Artymowicz1994} derived disk truncation radii of 0.2$a$-0.3$a$ for a
  reduced mass, $\mu = 0.3$ and an eccentricity, $e=0.4$, and this truncation radius decreases
  with increasing $\mu$. If we now consider the A and B components and adopt
  $M_{\rm A}=1.1\,M_\odot$\,\citep{Tokovinin1999}, $M_{\rm B}=M_{\rm Ba}+M_{\rm Bb}=1.3$\,$M_\odot$
  \citep{Boden2005}, $a_{A-B}=45$\,au, and $e=0.4$ \citep{Tokovinin2014}, the truncation
  radius of the outer disk is $\lesssim$10\,au. Modeling efforts by \citet{Pichardo2005}
  found that, for a system with $\mu\approx0.5$, the maximum value for the outer radius of
  the disk is 10\,\% of the semi-major axis, or $\sim$5\,au \citep[when using the updated
  semi-major axis of $a=1.03\,$arcsec in][]{Tokovinin2014}. Once again, this value is in
  perfect agreement with the VLA observations.

\end{enumerate}

Since the material in the disk cannot freely accrete onto the central sources or diffuse to
larger radii, its viscous evolution is considerably slowed down or stopped
completely. Yet, the extreme ultraviolet and X-ray emission from the Ba-Bb pair can
still produce mass loss via photoevaporative winds \citep{Font2004, Owen2011}. In
particular, \citet{Kastner2004} determined an X-ray luminosity value of
$L_{\rm X}=1.4\times10^{29}$\,erg\,s$^{-1}$ for HD~98800~B, which implies a mass-loss rate due
to photoevaporation of 5$\times$10$^{-10}$\,$M_\odot$\,yr$^{-1}$ using the disk wind
prescription in \citet{Owen2011}. As pointed out in \citet{Alexander2012}, such a mass
loss appears to be incompatible with the 3$\times$10$^{-4}$\,$M_\odot$ disk mass estimate in
\citet{Andrews2010a}, because the resulting disk lifetime ($\sim$ 0.5\,Myr) is much shorter
than the 10\,Myr age of the TW Hya association.  However, the presented VLA observations offer two
solutions to this problem by suggesting both a much higher mass value ($>5\,M_{\rm Jup}$ in our
MCFOST model) and a very small disk. The new mass yields an updated disk lifetime of
$>$10\,Myr by itself and a smaller disk size would decrease the mass-loss rate with respect
to a full disk given its smaller surface area, hence bringing the expected
photoevaporative rate to agreement with the age of the system.

All these results imply that the disk around HD~98800~B may have evolved considerably
slower than those around single stars due to the truncation of both its inner and outer
radii. This truncation produces an important (or total) suppression of accretion onto
Ba-Bb \citep[explaining the lack of accretion signatures,][]{Soderblom1996,
  Muzerolle2000}, and thus most of the mass remains locked in the disk and is subject to
evolution via photoevaporative winds. This scenario successfully explains the presence of a
massive, gas-rich disk in a 10\,Myr old multiple system, the apparent lack of accretion, as well as
possible free-free emission from this wind that could produce the much flatter spectral
index of the SED between the 8.8\,mm and 5\,cm VLA fluxes. 

Given that HD~98800 is a bound system, it is very likely that all of its components are
coeval. Moreover, the masses of A and B are somewhat comparable; \citet{Tokovinin1999}
proposed that they have the same stellar mass, and \citet{Prato2001} found values of 1.1 and
1.6\,$M_\odot$ for A and B, respectively. Therefore, the presence of a massive disk
around HD~98800~B and the diskless HD~98800~A (see also Sec.~\ref{sec:HD98800A} below) are
challenging if we assume that both formed with a disk \citep[e.g.,][]{Prato2001}.
The evolutionary scenario proposed earlier (i.e., the accretion rate is strongly suppressed in
HD~98800~B by Ba-Bb, and the disk is evolving through photoevaporative winds only) offers
an explanation to this issue. HD~98800~A is itself a binary (Aa-Ab), so it is possible
that accretion was also suppressed in it. However, \citet{Kastner2004} found its X-ray
luminosity to be $\approx$4\,times higher than in HD~98800~B. Assuming that this ratio has
been constant throughout the lifetime of the system, it implies a mass loss through
photoevaporative winds about $\approx$5 times higher for A
\citep[$\dot{M}_{\rm wind}\propto L_{\rm X}^{1.14}$,][]{Owen2011}, and thus its dispersal would
have been significantly faster. The mass of the disk is also known to scale linearly with
stellar mass \citep[although with significant scatter,][]{Andrews2013,Pascucci2016}, and
the results of \citet{Prato2001} then suggest that the disk around A was probably less
massive. These two facts combined could explain why HD~98800~A is already diskless.

Interestingly, the TW~Hya association contains another long-lived disk in a multiple
system with several features resembling those of HD~98800: TWA~3, a hierarchical triple
system that comprises a close binary \citep[Aa-Ab, e.g.][]{Muzerolle2000} surrounded by a
circumbinary disk and orbited by a visual companion at 40-110\,au \citep{delaReza1989,
  Reipurth1993, Webb1999, Torres2003, Kellogg2017}. The small binary separation and
eccentricity \citep[$a$=0.2\,au, $e$=0.6,][]{Kellogg2017} explain the 1\,au cavity
inferred from SED modeling \citep{Andrews2010a}. Such a small cavity may allow some mass
accretion via streams, accounting for its low accretion rates
\citep[5$\times$10$^{-11}$-2.5$\times$10$^{-10}$\,$M_\odot$\,yr$^{-1}$,][]{Muzerolle2000,
  Herczeg2009} and its observed periodicity, similar to the binary period
\citep{Tofflemire2017}.  As discussed in \citet{Kellogg2017}, this system is also
interesting from an evolutionary perspective, since the B component is diskless. The
evolutionary scenario proposed here for HD~98800 also works for TWA~3: while the disk
around TWA~3B likely evolved on regular timescales through viscous evolution and
photoevaporation, the mass-accretion rate of the TWA~3A is significantly reduced by the
Aa-Ab pair, resulting in a longer disk lifetime. Their ages, comparable disk properties,
and system architectures suggest that both HD~98800~B and TWA~3 could have survived
through the same mechanisms, and comparative studies of these two systems offer a great
opportunity for understanding the importance of star-disk interactions and multiplicity in
disk evolution.

Although the proposed scenario provides plausible explanations for several properties of
the disk around HD~98800~B, it also presents an issue: the simulations of
\citet{Artymowicz1996} predict that, despite the cavity formed in circumbinary disks by
the binary system, some material can still accrete onto the central sources through gas
streams. In the case of eccentric systems, this accretion would be periodic, with a period
similar to that of the binary. This phenomenon has been found in other circumbinary disks,
such as DQ~Tau \citep{Mathieu1997}, UZ~Tau~E \citep{Jensen2007}, and~TWA 3A\
\citep{Tofflemire2017}, which raises the question of why HD~98800~B does not show a
similar behavior. Although we do not have a clear explanation for this, a number of
scenarios could account for the apparent lack of periodic accretion in HD~98800~B. A
likely possibility is that periodic accretion actually exists in the system, but at a very
low rate that makes its detection challenging; in fact, \citet{Yang2012} suggest that
HD~98800~B is an accreting source. The authors note that previous studies were unable to
find accretion signatures in the system, and their low-resolution optical spectrum places
an upper limit of $10^{-9.9}$\,$M_\odot$ yr$^{-1}$ to its accretion rate. However, their
detection of H$_2$ suggests ongoing accretion at lower rates given that H$_2$ is not
detected in nonaccreting systems \citep{Ingleby2009} but can still be seen in very slow
accretors \citep{Ingleby2011}. It is even possible that the periodic nature of accretion
in the system is responsible for some of the previous nondetections of accretion
signatures. A second explanation could be that HD~98800~B displays more extreme orbital
parameters, because the three systems mentioned above have all shorter periods (and hence
smaller separations) and lower eccentricities: for comparison, DQ~Tau shows $P$=15.8\,days
and $e$=0.56\,\citep{Czekala2016}, UZ~Tau~E has $P$=19.2\,days and $e$=0.33
\citep{Jensen2007}, and TWA~3A has $P$=34.9\,days and $e$=0.63 \citep{Kellogg2017}, in
contrast with the $P$=315\,days and $e$=0.78 of HD~98800~B. It is thus possible that,
under these different orbital conditions, gas streams cannot form or are less
efficient. Moreover, the significantly longer period of HD~98800~B makes it harder to
study any periodic variability. Long-term monitoring with high sensitivity is required to
confirm the low mass-accretion rate of HD~98800~B and to identify any possible variability
in it.

\subsection{The emission from HD~98800~A}\label{sec:HD98800A}

HD~98800~A is detected at both 8.8\,mm and 5\,cm in the synthesized VLA images and becomes
brighter than B in the latter wavelength. The flux detected in both cases is well above
the expected photospheric level (see Figure~\ref{fig:SEDs}), indicating additional
contribution from other mechanisms. One possible explanation for this excess is the
presence of a previously unknown circumstellar disk around A. However, the emission is
unresolved at 8.8\,mm, meaning that the corresponding disk radius would have to be
$<2.5$\,au, smaller than the expected truncation radius due to the B
\citep[e.g.,][]{Papaloizou1977, Artymowicz1994}. The dust in such a disk would be very
close to the central stars and would thus emit significantly in the infrared. However, no
infrared excess has been found in HD~98800~A \citep[e.g.][]{Koerner2000, Prato2001}. Given
the negative spectral index of the emission and the partial polarization detected at
5\,cm, stellar activity appears as a better explanation the observations, as proposed by
\citet{Kastner2004} in the light of \emph{Chandra} X-ray observations of HD~98800; the A
component was found to be $\sim$4\,times brighter than B, and a flaring event was even
detected during the observations, suggesting that the star is significantly
active. Higher-resolution observations with ALMA will probe smaller scales and confirm or
deny the presence of a small disk around HD~98800~A, clarifying the nature of the
detected excess at mm/cm wavelengths.

%%%% CONCLUSIONS
\section{Summary and conclusions}\label{sec:conclusions}

We present Ka~band (8.8\,mm, 34\,GHz) and C~band (5\,cm, 6\,GHz) VLA observations of the
hierarchical quadruple system HD~98800 in the 10\,Myr old TW~Hya association. The 8.8\,mm
observations successfully resolve the disk and the inner cavity around the B component and
detect unresolved emission from A. Both A and B are also detected (unresolved) at
5\,cm. Only the 5\,cm emission from A shows signatures of polarization. We use the
MCFOST radiative transfer software to model both the SED and the resolved VLA image of the
disk. The main results of the study are:

\begin{enumerate}

\item The (sub)mm spectral index of the disk around HD~98800~B is $\approx2$ up to
  8.8\,mm, suggesting either optically thin radiation from a dust population with a gray
  opacity law at millimeter wavelengths or optically thick emission.

\item The disk extends from 3 to 5\,au at 8.8\,mm. The values of both the
  inner and outer radii (truncated by the Ba-Bb and the interaction with A, respectively)
  agree with model predictions of viscous disk interactions with companions.

\item For such a disk size, we were unable to reproduce the observed 8.8\,mm flux and
  spectral index with an optically thin disk. Combined with previous detections of gas in
  the system, this suggests that HD~98800~B is probably a massive, optically thick and
  gas-rich disk, more similar to protoplanetary than to debris disks.

\item Given that both disk fractions and exoplanet occurrence are lower in close binaries
  than around single stars, the survival of a massive disk in a quadruple system, such as
  HD~98800~B, appears to contradict disk evolution theories. We suggest that the inner and
  outer truncation of the disk have stopped the viscous evolution of the disk, which is
  thus governed by photoevaporation. The small size of the disk lowers the mass-loss rate
  through photoevaporation, explaining the longevity of the disk. This scenario results in
  longer disk lifetimes that may ease planet formation in multiple systems with the
  appropriate configuration.

\item The spectral index of HD~98800~B between 8.8\,mm and 5\,cm is significantly
  shallower ($\alpha=1.3$), indicating that additional contribution from free-free
  emission by photoevaporating material or from gyrosynchrotron emission from stellar
  activity is present at cm wavelengths. In case the photoevaporative wind is the correct
  explanation, then it is possible that the 8.8\,mm flux is also partially probing this
  emission, making HD~98800~B an ideal target to study this phenomenon.

\item The disk size is 2-3\,times smaller than the one derived by \citet{Andrews2010a},
  who used partially resolved SMA observations at 880\,$\mu$m. While we attribute the
  difference to the lower angular resolution of SMA, if this size dependence with
  wavelength is, in fact, real, it may hint at significant dust radial migration in the
  system.

\item The A component appears unresolved both at 8.8\,mm and 5\,cm, and the observed
  fluxes are above the expected photospheric levels. In contrast to HD~98800~B, the
  spectral index between the two VLA bands is negative ($\alpha=-0.4$). Together with the
  partial polarization of the 5\,cm emission of HD~98800~A, this favors a stellar origin
  for the excess instead of a previously unknown disk around this component.

\end{enumerate}

Follow-up ALMA observations will be able to search for CO in the system with increased
sensitivity with respect to previous studies and to measure the disk size at shorter
wavelengths to test the dust radial migration scenario in HD~98800~B. Additional VLA
observations at other wavelengths will also provide a better sampling of the mm/cm range
and will help to quantify the relative importance (if any) of free-free emission in the
disk. Given its proximity and unique properties, HD~98800 offers a unique laboratory to
study the impact of multiplicity on planet formation and disk evolution.

%%%% ACKNOWLEDGEMENTS
\vspace{1cm}

We thank the anonymous referee for their useful comments, which helped improving the
quality of this manuscript.  We thank Richard Alexander for discussion on disk evolution
in binary systems and Eric Nielsen for examining the orbit of HD~98800 including the VLA
data presented here. This manuscript is based on data from the NSF's Karl G. Jansky Very
Large Array (VLA). We have also used data from the \emph{Herschel Space Observatory}
(\emph{Herschel}). \emph{Herschel} is an ESA space observatory with science instruments
provided by European-led Principal Investigator consortia and with important participation
from NASA. 

\software{emcee \citep{emcee}, MCFOST \citep{MCFOST}, HIPE \citep[v15;][]{Ott2010}, CASA
  \citep{CASA}, Matplotlib  \citep{Matplotlib}, SciPy \citep{Scipy},
  Numpy \citep{Scipy}, pandas \citep{pandas}, Astropy \citep{Astropy2013}.}

\bibliography{biblio} 

\listofchanges

\end{document}